\documentclass[twocolumn]{aastex631}
\usepackage{CJK}
\usepackage{amsmath}
\usepackage{xcolor}
\usepackage{bm}
\usepackage{soul}
\shorttitle{Magnetic field spectral evolution in the inner heliosphere}
\shortauthors{Sioulas et al.}
\graphicspath{{./}{figures/}}

\begin{document}
\begin{CJK*}{UTF8}{gbsn}
\title{Magnetic field spectral evolution in the inner heliosphere}

\correspondingauthor{Nikos Sioulas}
\email{nsioulas@ucla.edu}

\author[0000-0002-1128-9685]{Nikos Sioulas}
\affiliation{Department of Earth, Planetary, and Space Sciences, University of California, Los Angeles \\
Los Angeles, CA 90095, USA}

\author[0000-0001-9570-5975]{Zesen Huang (黄泽森)}
\affiliation{Department of Earth, Planetary, and Space Sciences, University of California, Los Angeles \\
Los Angeles, CA 90095, USA}

\author[0000-0002-2582-7085]{Chen Shi (时辰)}
\affiliation{Department of Earth, Planetary, and Space Sciences, University of California, Los Angeles \\
Los Angeles, CA 90095, USA}

\author[0000-0002-2381-3106]{Marco Velli}
\affiliation{Department of Earth, Planetary, and Space Sciences, University of California, Los Angeles \\
Los Angeles, CA 90095, USA}

\author[0000-0003-2880-6084]{Anna Tenerani}
\affiliation{Department of Physics, The University of Texas at Austin, \\
     TX 78712, USA}

\author[0000-0002-4625-3332]{ Trevor A. Bowen}
\affiliation{Space Sciences Laboratory, University of California, \\ Berkeley, CA 94720-7450, USA}


\author[0000-0002-1989-3596]{Stuart D. Bale}
\affil{Physics Department, University of California, Berkeley, CA 94720-7300, USA}
\affil{Space Sciences Laboratory, University of California, Berkeley, CA 94720-7450, USA}

\author[0000-0002-9954-4707]{Jia Huang}
\affiliation{Climate and Space Sciences and Engineering, University of Michigan, \\  Ann Arbor, MI 48109,
USA}

\author[0000-0002-8700-4172]{Loukas Vlahos}
\affiliation{Deprtment of Physics,  Aristotle University of Thessaloniki\\
GR-52124 Thessaloniki, Greece }

\author[0000-0003-2845-4250]{ L. D. Woodham}
\affiliation{Department of Physics, The Blackett Laboratory, Imperial College London, London, SW7 2AZ, UK}

\author[0000-0002-7572-4690]{T. S. Horbury}
\affiliation{The Blackett Laboratory, Imperial College London, London, UK}

\author[0000-0002-4401-0943]{Thierry {Dudok de Wit}}
\affil{LPC2E, CNRS and University of Orl\'eans, Orl\'eans, France}

\author[0000-0001-5030-6030]{Davin Larson}
\affiliation{Space Sciences Laboratory, University of California, Berkeley, CA 94720-7450, USA}

\author[0000-0002-7077-930X]{Justin Kasper}
\affiliation{BWX Technologies, Inc., Washington DC 20002, USA}
\affiliation{Climate and Space Sciences and Engineering, University of Michigan, Ann Arbor, MI 48109, USA}

\author[0000-0002-5982-4667]{Christopher J. Owen}
\affiliation{Mullard Space Science Laboratory, University College London, Dorking, RH5 6NT, UK}

\author[0000-0002-7728-0085]{Michael L. Stevens}
\affiliation{Harvard-Smithsonian Center for Astrophysics, \\ Cambridge, MA 02138, USA}

\author[0000-0002-3520-4041]{Anthony Case}
\affiliation{Smithsonian Astrophysical Observatory, \\
Cambridge, MA 02138, US}

\author[0000-0002-1573-7457]{Marc Pulupa}
\affil{Space Sciences Laboratory, University of California, Berkeley, CA 94720-7450, USA}

\author[0000-0003-1191-1558]{David M. Malaspina}
\affiliation{Astrophysical and Planetary Sciences Department, University of Colorado, Boulder, CO, USA}
\affiliation{Laboratory for Atmospheric and Space Physics, University of Colorado, Boulder, CO, USA}

\author[0000-0002-0675-7907]{ J.W. Bonnell}
\affil{Space Sciences Laboratory, University of California, Berkeley, CA 94720-7450, USA}

··

\author[0000-0002-0396-0547]{Roberto Livi}
\affiliation{Space Sciences Laboratory, University of California, Berkeley, CA 94720-7450, USA}

\author[0000-0003-0420-3633]{Keith Goetz}
\affiliation{School of Physics and Astronomy, University of Minnesota, Minneapolis, MN 55455, USA}

\author[0000-0002-6938-0166]{Peter R. Harvey}
\affil{Space Sciences Laboratory, University of California, Berkeley, CA 94720-7450, USA}

\author[0000-0003-3112-4201]{Robert J. MacDowall}
\affil{Solar System Exploration Division, NASA/Goddard Space Flight Center, Greenbelt, MD, 20771}

\author[0000-0001-6172-5062]{Milan Maksimovi\'c}
\affiliation{LESIA, Observatoire de Paris, Universit e PSL, CNRS, Sorbonne Universit e, Universit e de Paris, 5 place Jules Janssen, 92195 Meudon, France}

\author[0000-0003-2783-0808]{P. Louarn}
\affiliation{
IRAP, Université Toulouse III - Paul Sabatier, CNRS, CNES, Toulouse, France}

\author[0000-0002-9975-0148]{A. Fedorov}
\affiliation{
IRAP, Université Toulouse III - Paul Sabatier, CNRS, CNES, Toulouse, France}







\begin{abstract}



Parker Solar Probe and Solar Orbiter data are used to investigate the radial evolution of magnetic turbulence between $0.06 ~ \lesssim R ~\lesssim 1$ au. The spectrum is studied as a function of scale, normalized to the ion inertial scale $d_{i}$. In the vicinity of the Sun, the inertial range is limited to a narrow range of scales and exhibits a power-law exponent of, $\alpha_{B} = -3/2$, independent of plasma parameters. The inertial range grows with distance, progressively extending to larger spatial scales, while steepening towards a $\alpha_{B} =-5/3$ scaling. It is observed that spectra for intervals with large magnetic energy excesses and low Alfv\'enic content steepen significantly with distance, in contrast to highly Alfv\'enic intervals that retain their near-Sun scaling. The occurrence of steeper spectra in slower wind streams may be attributed to the observed positive correlation between solar wind speed and Alfv\'enicity.

\end{abstract}

\keywords{Magnetohydrodynamics (1964), Solar wind (1534), Alfven waves (23)}


\section{Introduction} \label{sec:intro}

The solar wind flow transports a wide range of magnetic field and plasma fluctuations \citep{coleman_turbulence_1968,  velli_turbulent_1989}. Because fluctuations are predominantly Alfv\'enic (i.e., magnetic field and velocity fluctuations exhibit the correlations typical of outwardly propagating Alfv\'en waves) \citep{bruno_solar_2013}, and relative density fluctuations are very small solar wind turbulence is usually discussed within the phenomenologies of incompressible magnetohydrodynamic (MHD). 

During the expansion, non-linear interactions result in a cascade of the energy towards smaller scales \citep{ matthaeus_who_2011}. Therefore, the energy injected into the solar wind at large scales, likely of solar origin, cascades downwards until it reaches ion scales, at which point the dynamics involve kinetic processes and structures such as ion cyclotron damping, kinetic Alfv\'en waves, kinetic scale current sheets, etc. \citep{leamon_dissipation_1999, cranmer_ion_2001, dmitruk_test_2004,tenbarge_current_2013, karimabadi_coherent_2013}. Turbulence is thought to be one of the main processes contributing to the non-adiabatic expansion, as well as the acceleration of the solar wind (SW) \citep{matthaeus_who_2011}.
MHD turbulence phenomenologies predict different power law exponents depending on prevailing characteristics of turbulence, such as spatial wave-number anisotropy \citep{goldreich_toward_1995,goldreich_magnetohydrodynamic_1997}, intermittency \citep{Pouquet_steepening_intermittency, Chandran_2015}, and the scale-dependent correlation between velocity and magnetic field  \citep{PhysRevLett.96.115002,Beresnyak_2010}. The variability of solar wind turbulence properties in the inner heliosphere reflects the diversity of solar coronal sources, that modulate the density, velocity, temperature, and ion composition of the plasma. As a result, several factors, including the role played by large-scale gradients \citep{velli_turbulent_1989, Chandran_perez}; the proximity to the heliospheric current sheet \citep{chen_near-sun_2021,shi_influence_2022}; the presence of magnetic field switchbacks \citep{martinovic_multiscale_2021,bourouaine_turbulence_2020, Shi_2022_patches}; large-scale velocity shear in the SW \citep{coleman_turbulence_1968}, strongly influence the properties of turbulence, resulting in a wide range of spectral scalings. By means of fitting the power-spectrum within a constant range in the frequency domain, recent statistical studies of PSP data, have recovered a non-evolving velocity spectral index close to $-3/2$, independent of the radial distance from the Sun \citep{shi_alfvenic_2021}, while the magnetic field spectrum steepens from a -$3/2$ slope at $\sim 0.2 \ au$ to a -$5/3$ slope at $\sim 0.6 \ au$  \citep{chen_evolution_2020, shi_alfvenic_2021}. 

Two scales are crucial to understanding the radial evolution of turbulence in the solar wind (1) the ion inertial scale $ d_{i} = { V_{A}}/{\Omega_{i}}$, and (2) the thermal ion gyroradius, $ \rho_{i} = { V_{th,i}}/{\Omega_{i}}$, where, $\Omega_{i} = {e|B|}/{m_{p}}$, is the proton gyrofrequency, $e$ is the elementary charge, $|B|$ is the magnitude of the magnetic field, and $m_{p}$ is the mass of the proton. With increasing heliocentric distance, both physical scales ($d_i$, $\rho_i$) increase \citep{Duan_spectral_break_high_freq, cuesta_intermittency_2022}. 

It is thus natural to expect that the relative physical scale of fluctuations of a given frequency decreases as the solar wind expands. Here we aim to understand the radial evolution of magnetic turbulence and to study the basic features of scaling laws for solar wind fluctuations in terms of properly normalized physical scales. High resolution data from Parker Solar Probe ($PSP$) \citep{fox_solar_2016}, and Solar Orbiter ($SO$) \citep{muller_solar_2020} covering heliocentric distances $13 \ R_{\odot}\lesssim R\lesssim 220$~$R_{\odot}$ are utilized and the radial evolution of the magnetic spectral index as a function of normalized wavenumber is investigated.

It is shown that closer to the Sun the magnetic field power-spectrum exhibits a poorly developed inertial range that is characterized by a $-3/2$ spectral index. The inertial range extends to larger and larger scales as the solar wind expands into the interplanetary medium, with the inertial range spectral index steepening towards a $-5/3$ value. We demonstrate, that the rate at which the steepening occurs is strongly dependent on magnetic energy excess and Alfv\'enicity of the fluctuations.


\begin{figure}[htb!]

\includegraphics[width=0.47\textwidth]{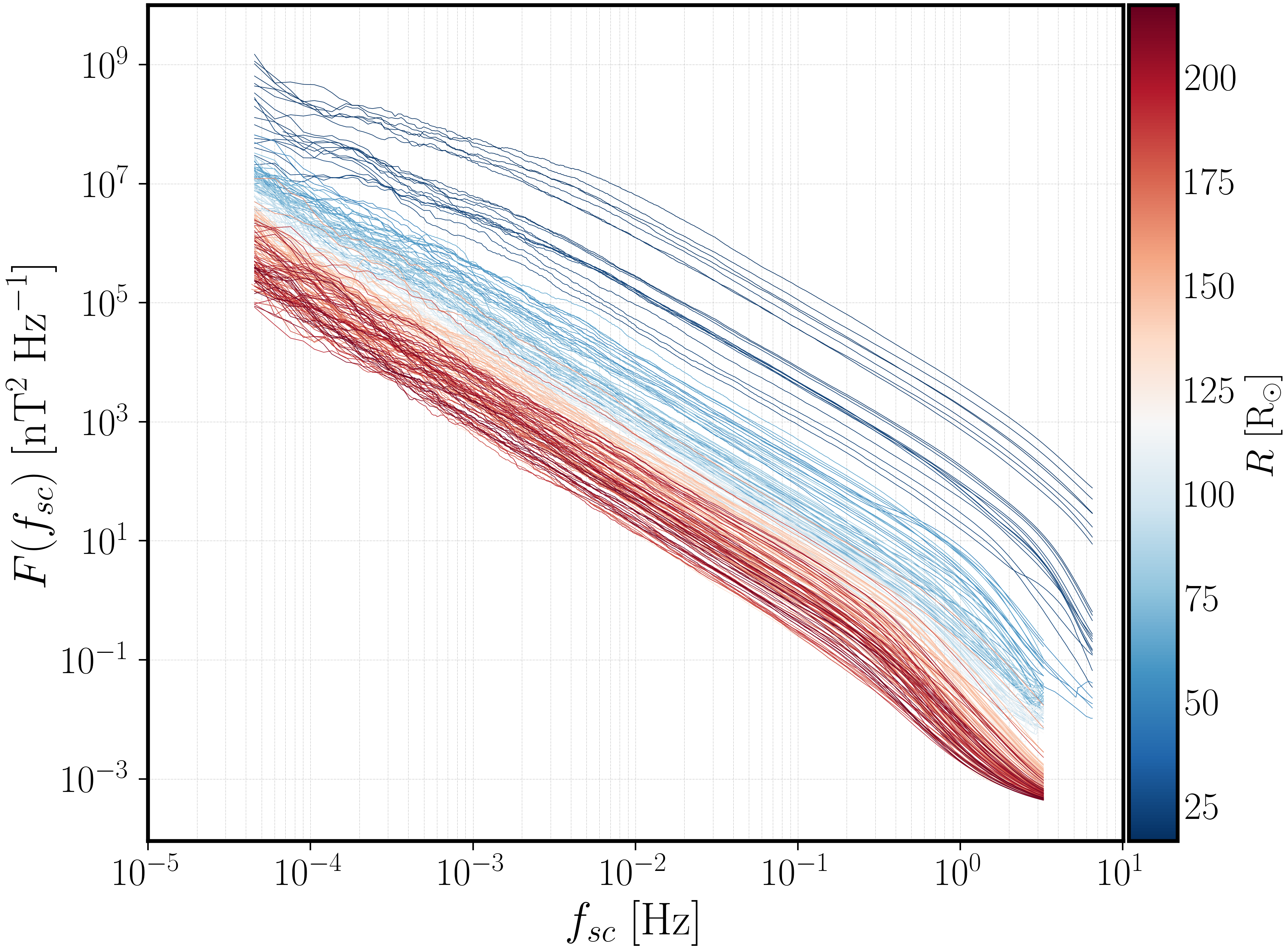}
\includegraphics[width=0.47\textwidth]{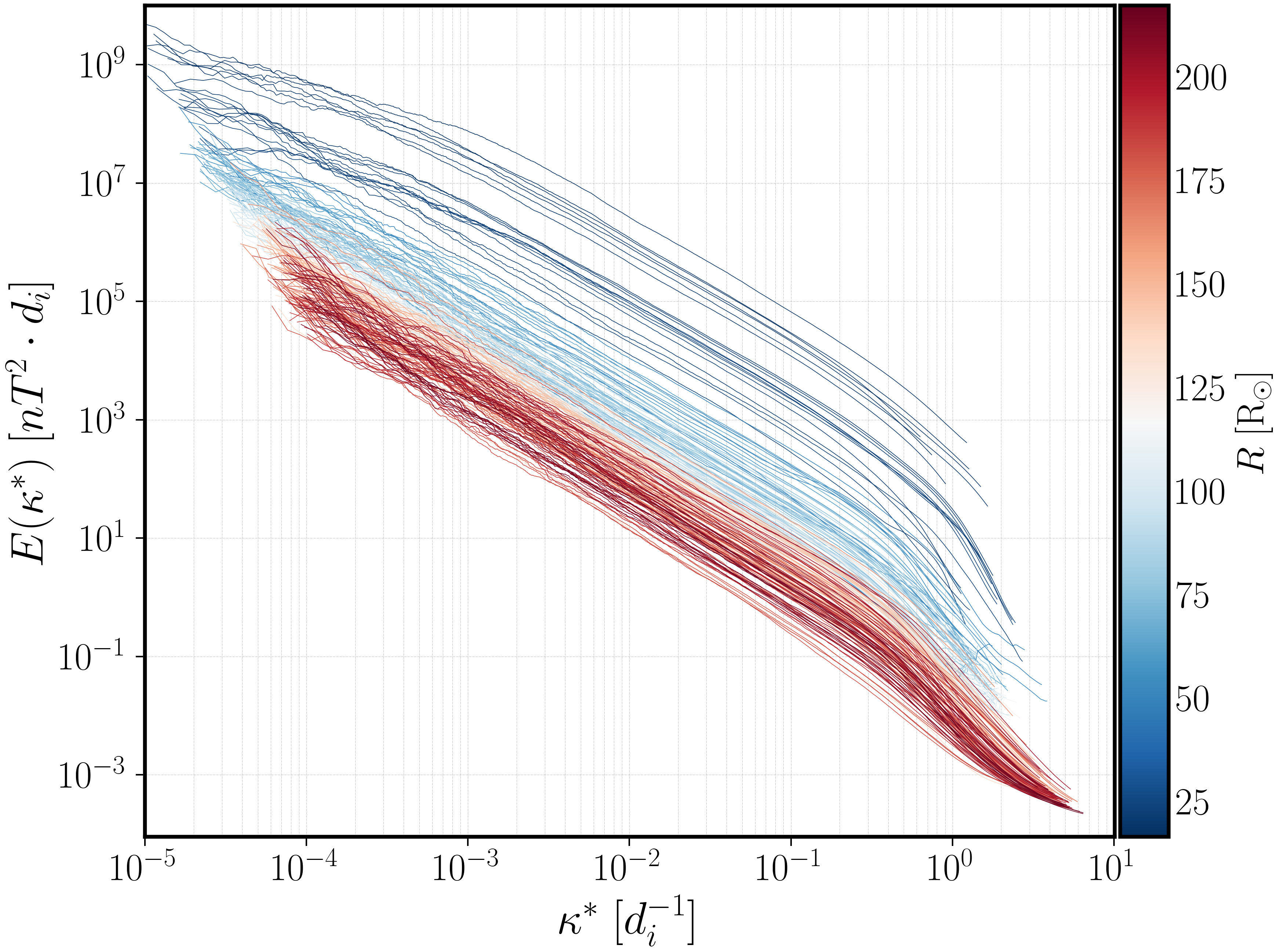}
\caption{ Magnetic field power spectrum, $PSD$ at different heliocentric distances. The power-spectrum is shown, as a function of (a) spacecraft frequency; (b) wavenumber $k^{∗} = \ell^{-1}$ in units of $d_{i}$
}\label{fig:PSD_B}
\end{figure}

\section{Radial Evolution of magnetic field spectral index}\label{sec:Results_mag_index}

We considered overlapping intervals of duration $d = 24$ hours such that the beginnings of adjacent intervals are $8$ hours apart. For each interval, the Fourier trace power spectral density $F(f_{sc})$ was calculated, smoothed by averaging over a sliding window of a factor of 2, and transformed into a wavenumber spectrum expressed in physical units $E(\kappa^{\ast})$ by virtue of the modified TH: 

\begin{equation}
   E(\kappa^{\ast}) = \frac{V_{tot}}{2 \pi \cdot \xi} F(f_{sc}) \  [nT^{2} \cdot \xi],
\end{equation}

\noindent where $\kappa^{\ast} = \kappa \cdot \xi = \frac{2 \pi f_{sc} }{V_{tot} }\cdot \xi$, and, $\xi = d_{i}, \rho_{ci}$.

 The radial evolution of the PSD ad a function of spacecraft frequency  and normalized  by $d_{i}$ is presented In Fig. \ref{fig:PSD_B}a,b respectively. Due to the expansion of the solar wind but also in part because of the turbulent cascade, a decrease of $\sim 4$ orders of magnitude in magnetic power is observed with increasing heliocentric distance. \par
\begin{figure}[htb!]
\includegraphics[width=0.47\textwidth]{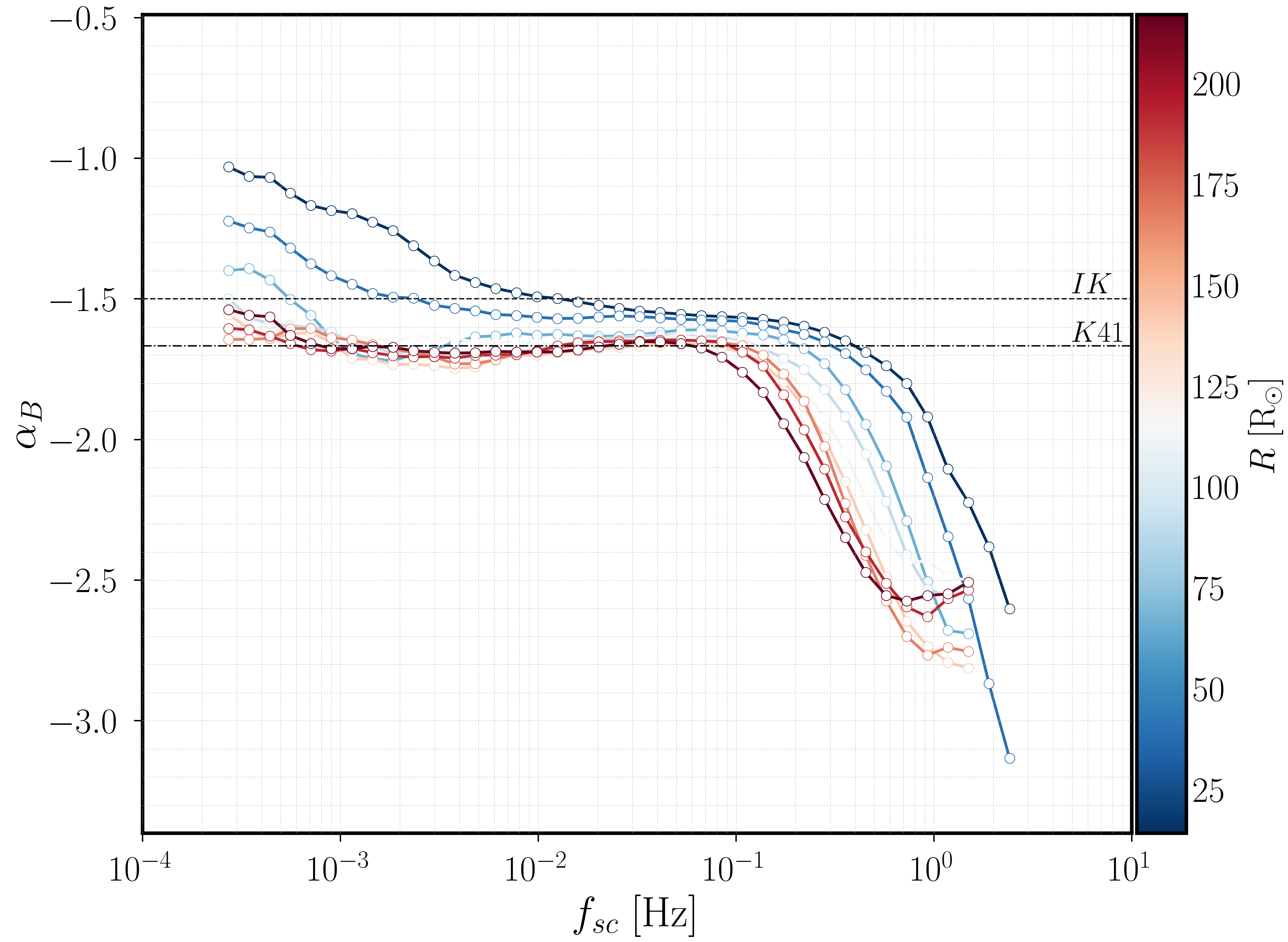}
\includegraphics[width=0.47\textwidth]{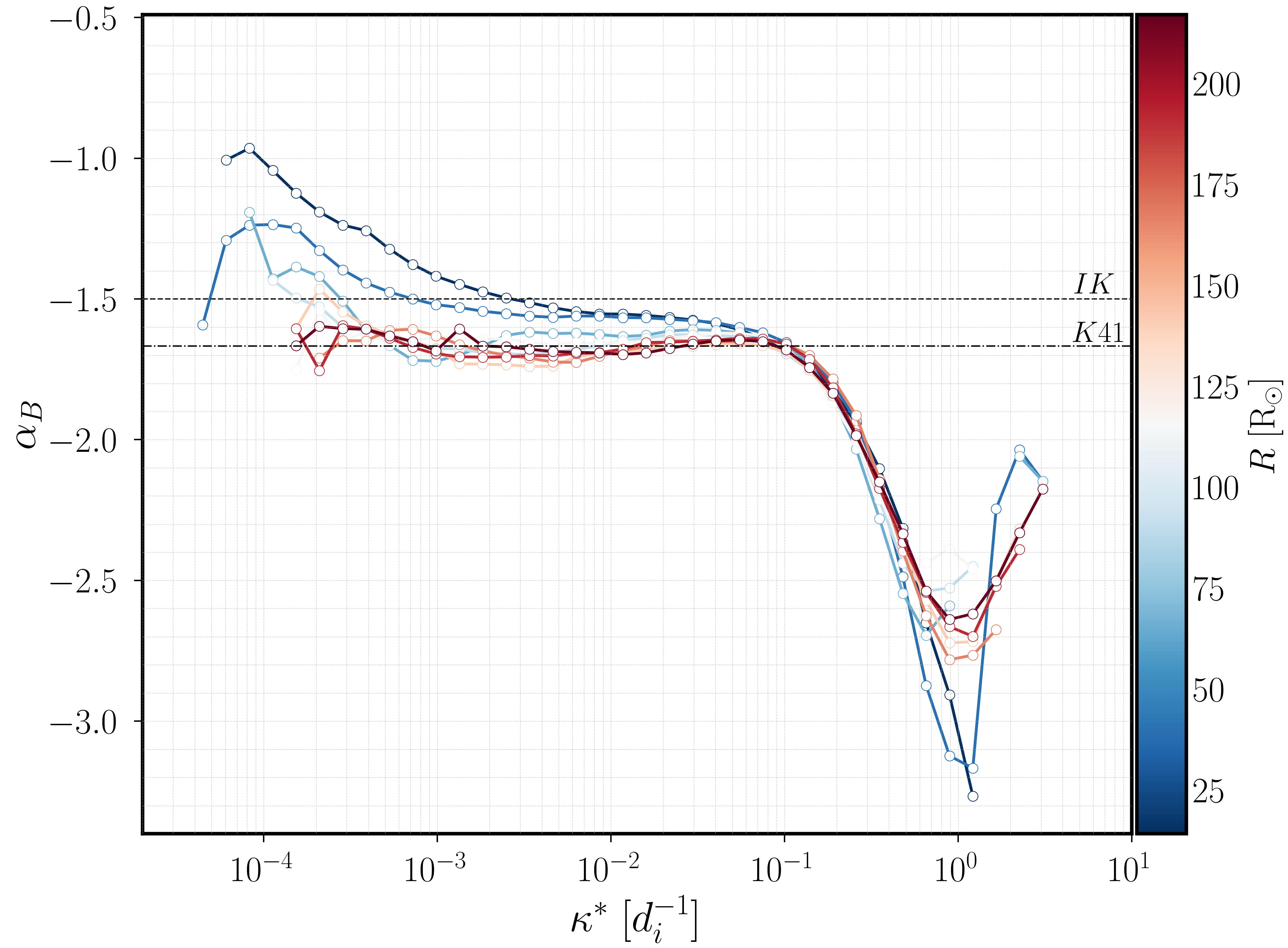}
\caption{ Evolution of magnetic field spectral index ($\alpha_{B}$) as a function of distance, \&  (a) frequency in units of $Hz$ (b)  normalized wavenumber $\kappa^{\ast}$ in units of $d_{i}$.}
\label{fig:ab_dist_time}
\end{figure}

The spectral index, $\alpha_{B}$ is obtained by taking a sliding window of one decade in the spacecraft-frame frequency (wavenumber) domain, over the smoothed spectra and calculating the best-fit linear gradient in log-log space over this window. For clarity, ten radial bins have been used, and the median value of the spectral index as a function of frequency has been estimated for intervals that fall within the same bin. The color of the curve is keyed to the mean value of the distance $R$ corresponding to the intervals within each bin. The results of this analysis are presented in Figure \ref{fig:ab_dist_time}a. In the inertial range, an energy cascade rate that is independent of scale is expected, reflecting on the power-spectrum in the form of a constant spectral index over this range of scales. In light of this, it can be seen that close to the Sun (dark blue line in Figure \ref{fig:ab_dist_time}a), the inertial range is limited into a narrow range of frequencies ($2 \times 10^{-2} - 2 \times 10^{-1} $Hz). As the solar wind expands in the interplanetary medium (1) a universal steepening (i.e., across all frequencies) is observed for the spectral index, $\alpha_{B}$, at a constant $f_{sc}$; (2) The curves shift horizontally to lower and lower frequencies. As illustrated in Figure \ref{fig:ab_dist_time}a, the frequency range over which the spectral index is constant is migrating to the left while steepening with increasing distance, from $\alpha_{B} \approx -3/2$ to $\alpha_{B} \approx -5/3$. Similar behavior is observed at the largest scales. Closer to the Sun for $f_{sc}\leq 2 \times 10^{-2}$ Hz, the spectrum gets progressively shallower at lower frequencies and obtains a value of $\alpha_{B} \approx -1$  at $f_{SC} =  \ 3 \times 10^{-4}$ Hz. As heliocentric distance increases, this low-frequency part of the spectrum gradually steepens, with all the frequencies approaching a $-5/3$ scaling. Therefore, as the solar wind propagates outward, the inertial range of the spectrum develops gradually, extending from higher frequencies to progressively lower and lower frequencies. Additionally, in accordance with \citep{Duan_spectral_break_high_freq} the ion scale break, separating the inertial from the kinetic range is observed to migrate to lower frequencies with distance. \par 
To cast the results in terms of relevant physical scales, we considered the evolution of $\alpha_{B}$ into the wavenumber domain normalizing by either the ion inertial length ($d_{i}$) or the ion gyroradius ($\rho_{i}$).

The evolution of the spectral index as a function of distance ($R$) in the wavenumber domain normalized by $d_{i}$, is illustrated in Figure \ref{fig:ab_dist_time}b. It is readily seen, that the vertical shifting of the curves to lower frequencies, observed in Figure \ref{fig:ab_dist_time}a, has vanished: all the curves  roll over at $\kappa d_{i}\approx~ 0.1$ and overlap at smaller scales. The normalization does not appear to have a substantial impact on the radial development of the spectral index at large scales, $\kappa^{\ast} \lesssim 8 \times 10^{-2}$, since a steepening closely resembles Figure \ref{fig:PSD_B}a is obtained. On the other hand, as shown in Figure \ref{fig:PSD_B}b, the small scale break, demarcating the beginning of the transition region, $\kappa^{\ast} \approx 9 \times 10^{-2} \ (\rho^{-1}_{i})$, does not show any remarkable evolution with distance and stays constant in physical space. We do not show plots using $\rho_{i}$ as normalization because the spectra do not collapse as clearly into one curve at small scales, demonstrating that $d_{i}$ is the
 more appropriate scale for such a normalization.


\subsection{Dependence of $\alpha_{B}$ on plasma parameters}\label{sec:Results_spectrum}

To disentangle the spectral variation with distance from changes due to the differing plasma parameters of different solar wind streams the dependence of $\alpha_{B}$ on the normalized cross helicity $\sigma_{c}$  \begin{equation}
    \sigma_{c} = \frac{E_{o} - E_{i}}{E_{o} + E_{i}},
\end{equation}
a measure of the relative amplitudes of  inwardly and outwardly propagating Alfvén waves, and the normalized residual energy $\sigma_r$
 \begin{equation}
    \sigma_{r} = \frac{E_{V} - E_{b}}{E_{V} + E_{b}},
\end{equation}
indicating the balance between kinetic and magnetic energy is examined. $E_{q} = \frac{1}{2}\langle \delta \boldsymbol{q}^{2} \rangle$ denotes the energy associated with the fluctuations  of the field $\boldsymbol{q}$. 
In particular, $E_{o,~ i}$ can be estimated using Elsasser variables, defining outward and inward propagating Alfvénic fluctuations \citep{Velli_91_waves, Velli_93} 
\begin{equation}
    \delta \boldsymbol{Z}_{o,i} = \delta\boldsymbol{V} \mp sign(B^{R}_{0})\delta \boldsymbol{b}, 
\end{equation}
$\delta \boldsymbol{B}  = \boldsymbol{B} - \boldsymbol{B_{0}}$, $\boldsymbol{B_{0}}$  the background magnetic field, $\delta \boldsymbol{b} = {\delta \boldsymbol{B}}/{\sqrt{\mu_0 m_p n_p}}$ the magnetic fluctuations in Alfv\'en units and $B^{r}_{0}$ the ensemble average of $B_{R}$, utilized to determine the polarity of the radial magnetic field  \citep{shi_alfvenic_2021}. The variation of $\alpha_{B}$ with
$V_{sw}$, the ratio of magnetic to thermal pressure, $\beta \equiv n_{p} K_{B} T / (B^{2}/2 \mu_{0}) \ll 1$, and the field/flow angle $\Theta_{BV}$ was also examined. Though we do not focus on $\beta$, and $\Theta_{BV}$ here, we will comment on these in Section \ref{sec:Summary}. The evolution of $\alpha_{B}$ is investigated by fitting the magnetic spectrum over a constant range ($10^{-3}- 3 \times 10^{-2}  \ d_{i}^{-1}$). To ensure that the plasma parameters under study do not vary significantly within the interval the duration of intervals has been reduced to $d=1$ hr.

\begin{figure*}
\centering
\setlength\fboxsep{0pt}
\setlength\fboxrule{0.0pt}
\fbox{\includegraphics[width=0.8\textwidth]{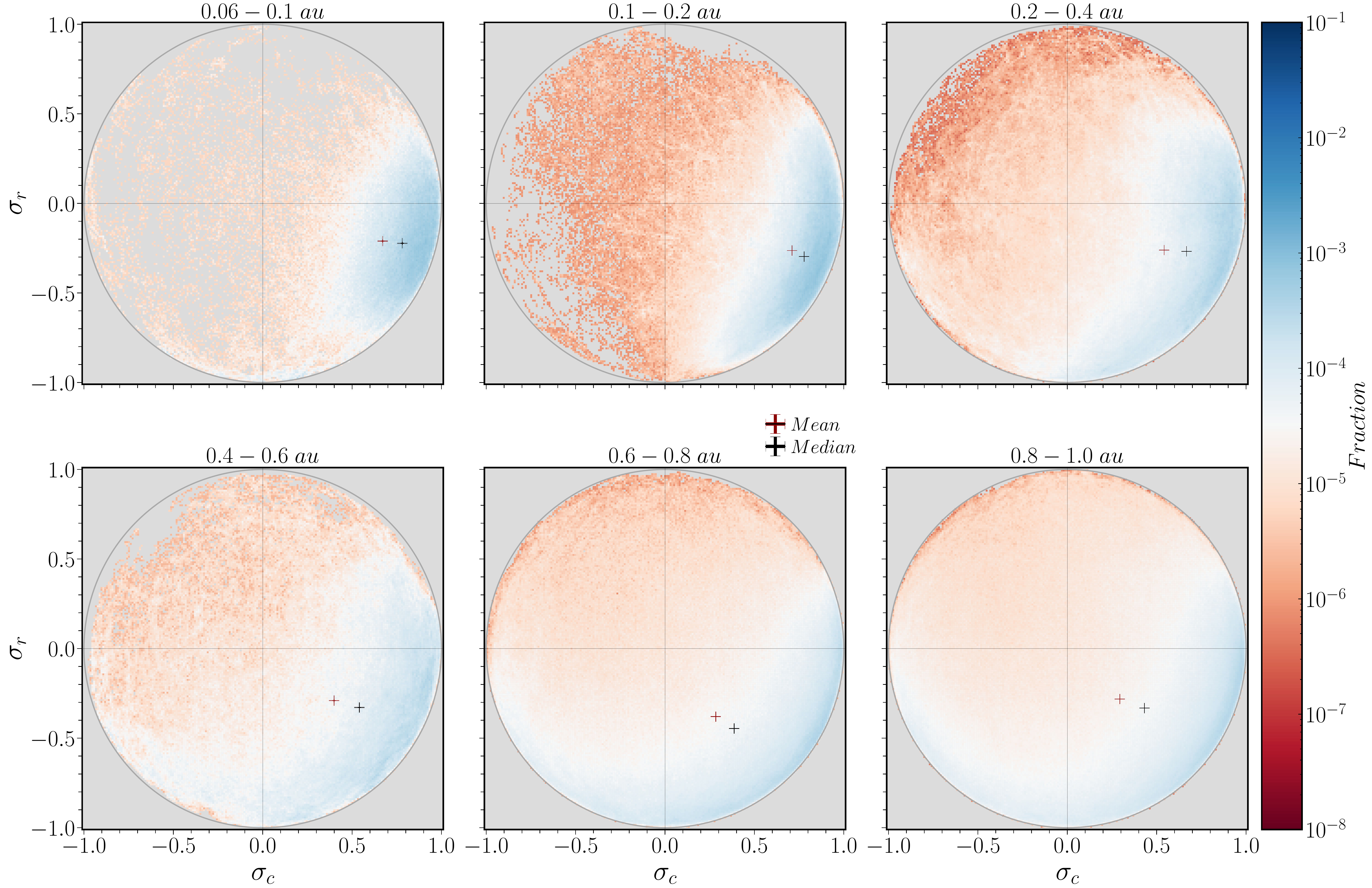}}
\caption{ The joint distribution of normalized cross-helicity $\sigma_{c}$ and normalized residual energy $\sigma_{r}$ at different heliocentric distances.}
\label{fig:sigma_c_sigma_r_joint}
\end{figure*}

\subsection{Solar Wind Speed, $V_{SW}$}\label{subsubsec:solar wind speed}

As shown in  Figure \ref{fig:sigma_c_sigma_r_spectral_index}a, Within 30 $R_s$, no significant differences in spectral index with solar wind speed are found with an inertial range scaling , $\alpha_{B} \approx -3/2$. As the solar wind expands, the dependence on solar wind speed becomes more evident: steepening occurs regardless of solar wind speed, but it is more efficient for slower solar wind streams. As a result, at $R \approx 1$ au, the dependence of the spectral index on speed is clear, with the spectral index being consistent with a K41 scaling in the fast wind and a steeper scaling of $\approx -1.8$ for the slowest winds. Categorizing the spectral index as a function of $\tau_{adv}$, Figure \ref{fig:sigma_c_sigma_r_spectral_index}b, instead of radial distance one finds that for $\tau_{adv} < 40 Hrs$ no clear trend is observed for the spectral index as a function of wind speed. Beyond, this point, though steepening is monotonic with $\tau_{adv}$ at all wind speeds. Overall, lower speed intervals display a significant radial steepening as compared to faster winds that only display a slight steepening. Closer to the Sun, however, there seems to be no dependence on wind speed on the spectral index, suggesting that the spectra are initially similar regardless of the wind speed.







\subsection{Normalized Cross Helicity, $\sigma_{c}$, $\&$ Normalized Residual Energy $\sigma_{r}$}\label{subsubsec:sigma_c}

The joint $\sigma_{c}$~-~$\sigma_{r}$ distribution, estimated using 1 minute-long moving averages of the respective timeseries is presented in Figure \ref{fig:sigma_c_sigma_r_joint}. The median and mean value of $\sigma_c$ and $\sigma_r$ for each bin are also shown as red and black crosses respectively. The gray circle  defines fluctuations with perfect alignment between velocity and magnetic field, given by $\sigma_{c}^{2} + \sigma_{r}^{2} =1$. Closer to the sun (0.06-0.1 au) turbulence is highly Alfv\'enic, dominated by outwardly propagating waves ($\sigma_c \approx 0.85 $), and in slight excess of magnetic energy ($\sigma_r \approx - 0.15$). A small population of strongly magnetically dominated intervals characterised by very low alfv\'enic content (i.e., $\sigma_r\approx -1$, and $\sigma_c\approx 0$, mostly associated with heliospheric current sheet (HCS) crossings is also observed (see \citep{shi_influence_2022}). At larger heliocentric distances the mean/median value of $\sigma_c$ progressively decreases \citep{chen_evolution_2020,shi_alfvenic_2021}. Several mechanisms have been proposed to explain the diminishing dominance of outwardly propagating waves with increasing heliocentric distance due to wave reflection, including velocity shears  \citep{1982JGR....87.3617B} and the parametric decay instability \citep{Tenerani_Velli_Param_decay, Shoda_2019}.  At 1au, $\sigma_{r}$ is clearly more negative than in the near-Sun environment, but it does not show a clear trend with radial distance. In the distance range of 0.6-1 au, most of the data points are concentrated in the lower half, with a few intervals having slightly positive $\sigma_r$ values. In addition, datapoints located in the bottom left quadrant are increasing with distance, indicating a radially decreasing dominance of waves propagating outward \par 

\begin{figure*}[htb!]
\centering
\setlength\fboxsep{0pt}
\setlength\fboxrule{0.0pt}

\includegraphics[width=0.4\textwidth]{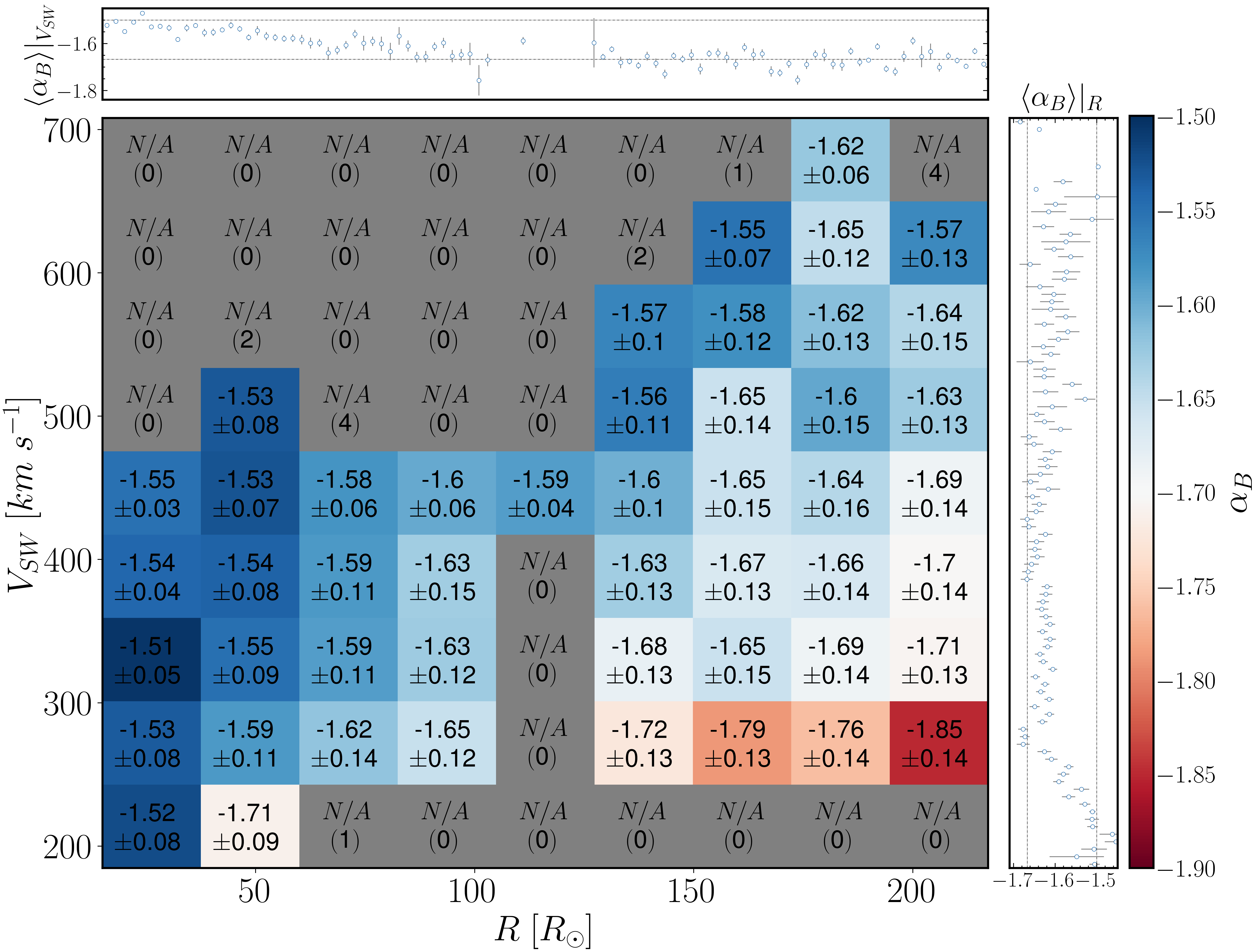}
\includegraphics[width=0.4\textwidth]{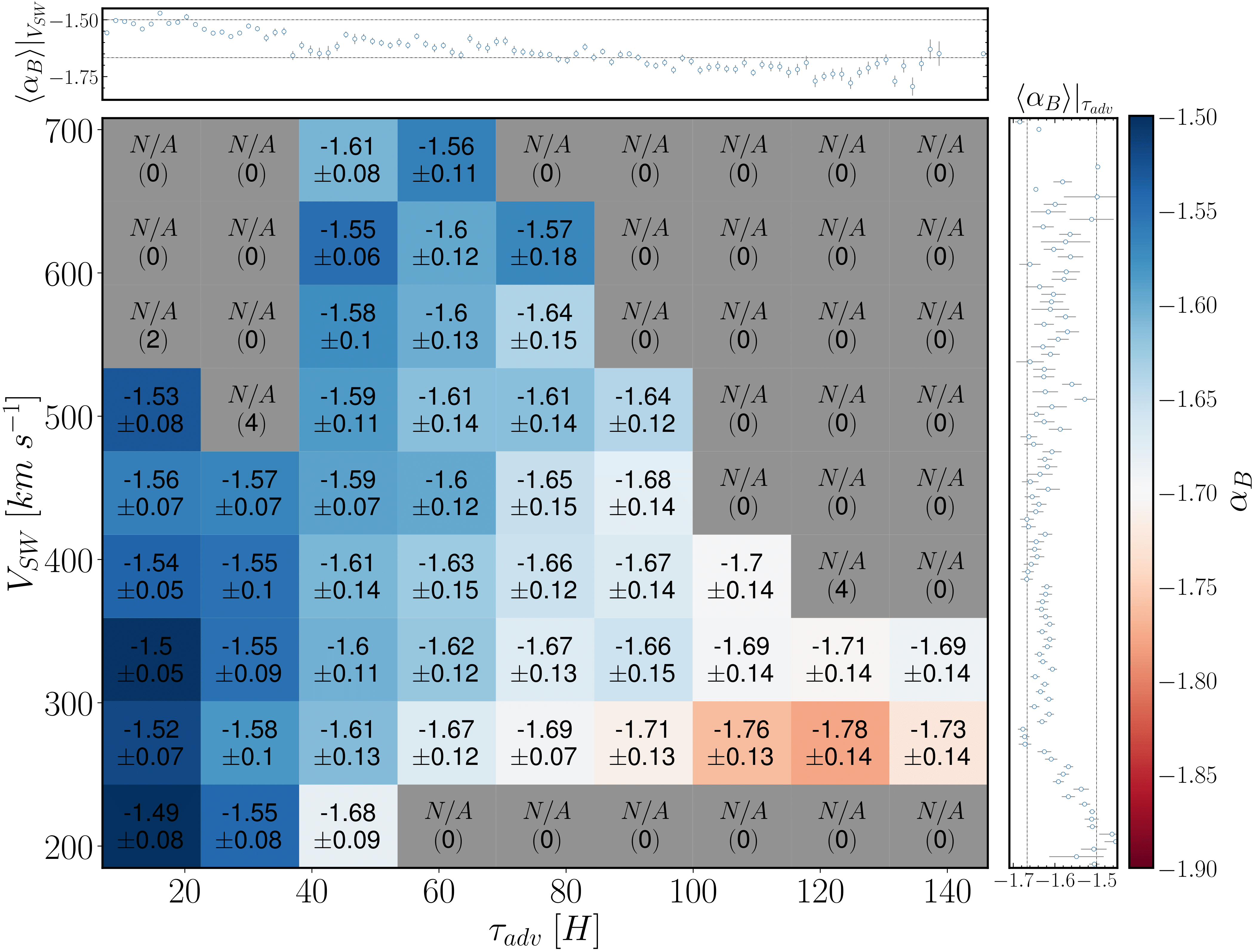}
\includegraphics[width=0.4\textwidth]{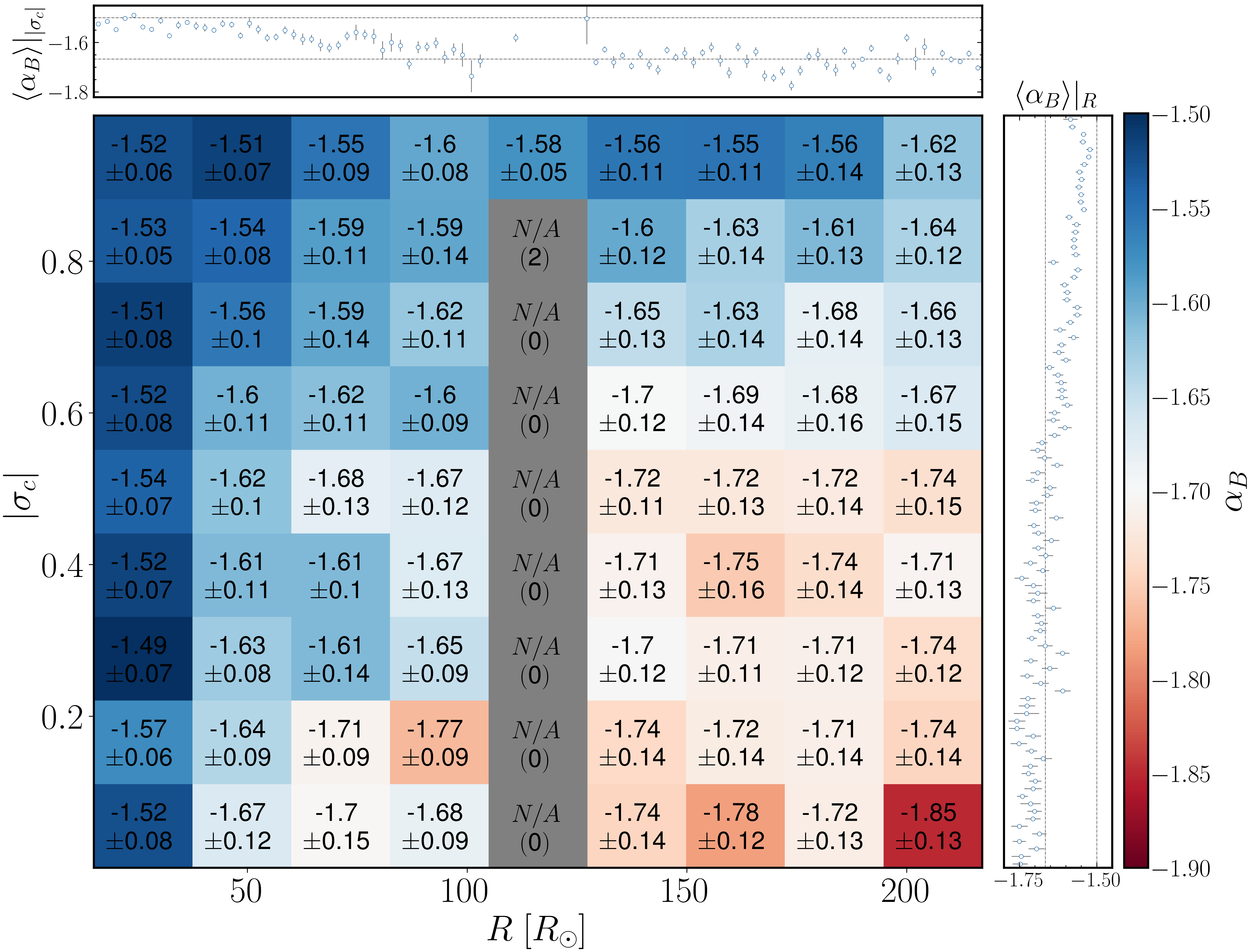}
\includegraphics[width=0.4\textwidth]{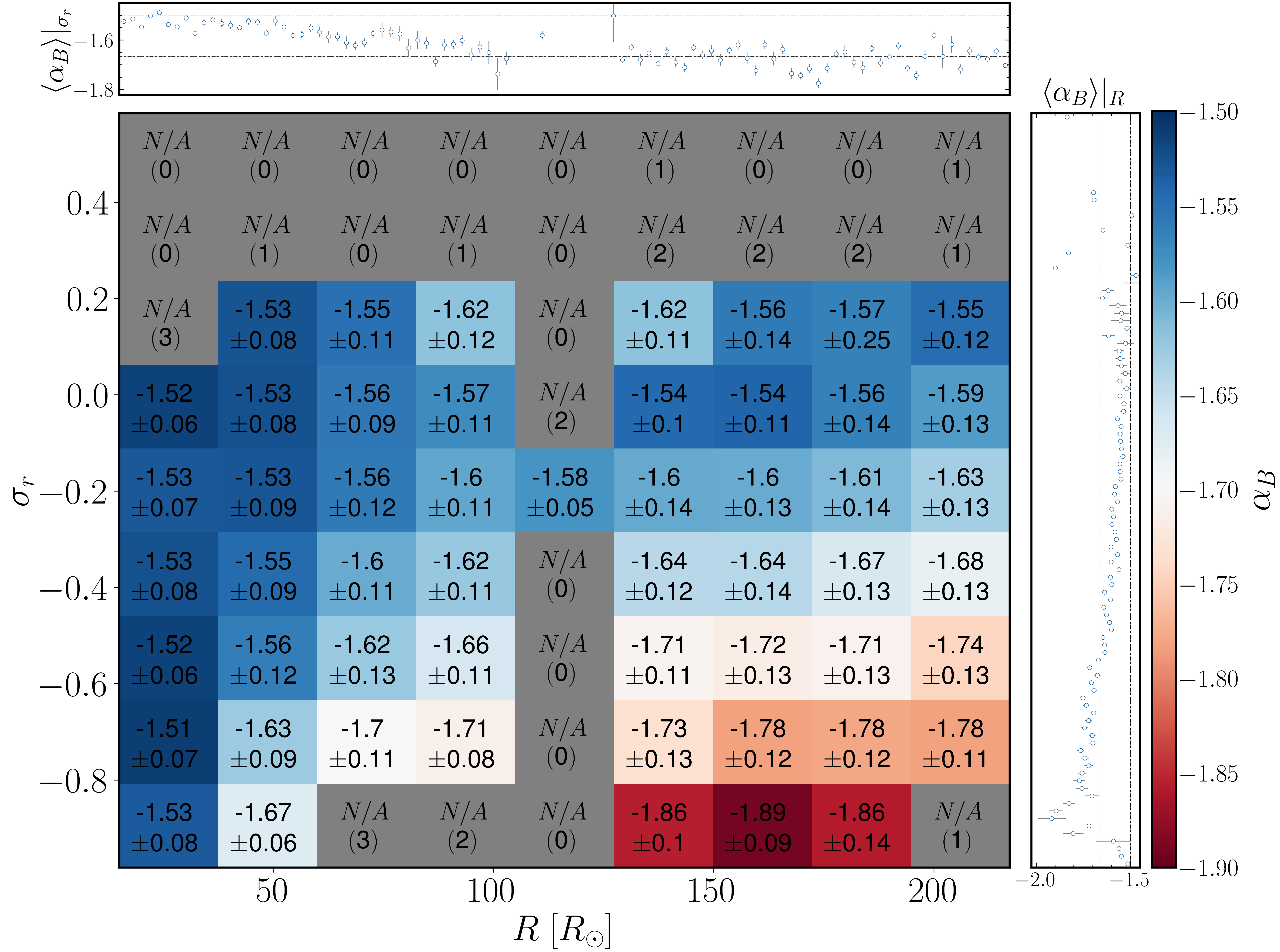}
\caption{Magnetic field spectral index $a_{B}$ as a function of  $V_{sw}$ and  (a) heliocentric distance, (b)  advection time $\tau_{adv},$  as well as, a function of distance and  (c) normalized cross helicity ($\sigma_c$), (d) normalized residual energy ($\sigma_r$)}\label{fig:sigma_c_sigma_r_spectral_index}
\end{figure*}

The power-spectra of the fluctuating fields $\delta \boldsymbol{b}, \delta \boldsymbol{V},\delta \boldsymbol{Z}_{o,i}$ have been obtained and both $\sigma_{c}$, and $\sigma_{r}$ have been estimated by integrating the resulting spectra over a constant range ($10^{-3}- 5 \times 10^{-2}  \ d_{i}^{-1}$ ) in the wavenumber domain normalized by the ion inertial length. The dependence of the spectral index on $|\sigma_{c}|$ and $\sigma_{r}$ as well as the radial distance (R) is presented in Figure \ref{fig:sigma_c_sigma_r_spectral_index}c,d for $\sigma_c$ and $\sigma_r$ respectively. These show how highly alfv\'enic ($|\sigma_{c}| \approx 1$) and energetically equipartitioned intervals display little spectral evolution, while evolution to significantly steeper spectra is associated with low $|\sigma_{c}|$ and/or large magnetic energy excess, with the data at large distances consistent with 1 AU results \citep{Podesta_borovsky_sigma_c, Chen_1u_residual_energy, Bowen_residual_energy}.

\section{Conclusions}\label{sec:Summary}

Using $PSP$ and $SO$ data from the inner heliosphere we have analyzed 1) how the statistical signatures of turbulence evolve with heliocentric distance and (2) the plasma parameters driving the evolution.

Identifying a plasma scale that grows radially  at the same rate as the high frequency break point is crucial for this study in order to anchor the spectrum in normalized wavenumber space and enable meaningful comparison between intervals sampled at different heliocentric distances. Since the high frequency break point exhibits a power-law  radial dependence with a scaling exponent $ 1.08 \pm 0.03$ \citep{Duan_spectral_break_high_freq, Lotz_high_freq_break}, then  $d_{i}$ which grows radially as $\propto R^{1.04 \pm 0.01}$ provides a better normalization than $\rho_{i}$ which is characterized by a $\propto R^{1.48 \pm 0.02}$ radial scaling.  It is important to emphasize that the goal here is not to find the quantity that has 1:1 correspondence with the high frequency break point. In fact, it is well known that intervals characterized by low ion $\beta$ values exhibit a magnetic power spectral density that breaks at the ion inertial length ($d_i$), while high $\beta$ intervals are characterized by a small scale break at the thermal ion gyroradius ($\rho_i$) \citep{2014GeoRL..41.8081C}.  Thus, the pinning of the power-spectrum at a constant $\kappa d_i$ scale, with increasing distance may be explained by the fact that plasma $\beta$ values remain rather low in the inner heliosphere, see Figure \ref{fig:overview}. Another plasma scale that is correlated with the high frequency break point and should be considered by future investigations is the proton cyclotron resonance \citep{Woodham_2018}.

Additionally, our analysis indicates that closer to the Sun, the inertial range of the magnetic field power-spectrum is poorly developed i.e., the range of scales over which $\alpha_{B}$ remains constants is limited; its value 
is closer to  $\alpha_{B} = -3/2$. As the solar wind expands into the interplanetary medium, the inertial range extends to progressively larger scales, while at the same time the inertial range spectral index steepens towards $\alpha_{B} = -5/3$.

We demonstrate that the rate at which $\alpha_{B}$ steepens is strongly dependent on the normalized residual energy and normalized cross helicity of the intervals under study. In particular, intervals with high alfv\'enic content ($|\sigma_{c}| \approx ~1 $), and equipartitioned in $E_{V}$-$E_{b}$ ($\sigma_{r} \approx ~0 $) seem to retain their near-Sun scaling, and show a minor steepening with radial distance. In contrast, magnetically dominated and balanced intervals are observed to strongly steepen, resulting in anomalously steep inertial range slopes at 1 au, consistent with previous studies \citep{Podesta_borovsky_sigma_c, Chen_1u_residual_energy, Bowen_residual_energy}.

While $|\sigma_{c}|\approx$ 1   and  $\sigma_{r} \approx 0$ values may be found in slow wind streams, especially closer to the sun, they are statistically less relevant than in fast winds \citep{shi_alfvenic_2021}. As a result, the occurrence of steeper spectral indices in slower wind streams may be attributed to the observed positive correlation between solar wind speed and $\sigma_{c}, ~ \sigma_{r}$.

\vspace*{0.1cm}
\par

Intervals with large magnetic energy excess closer to the Sun do not display the steep spectra observed at 1 au, attributed by \cite{Bowen_residual_energy} to the correlation between magnetic coherent structures and highly negative $\sigma_{r}$ values \cite{Pouquet_steepening_intermittency}.
Recent studies \citep{cuesta_intermittency_2022, Sioulas_2022}, suggest that magnetic field intermittency is strengthened with increasing heliocentric distance in the inner heliosphere, but no similar analysis has been conducted for the velocity field. However, velocity spectra do not display radial evolution \citep{shi_alfvenic_2021} and exhibit a scaling of $a_{v} = -3/2$ at 1 au \citep{Chen_1u_residual_energy}. Based on our results, we expect that both the magnetic and velocity field spectra display a $-3/2$ scaling closer to the Sun, with the evolution of the magnetic spectrum related to the in-situ generation of magnetic coherent structures during expansion. A study of the evolution of $\alpha_{B}$ and $a_{v}$ as a function of radial distance as well as intermittency is ongoing.
Turbulence in the solar wind is anisotropic with respect to the mean magnetic field  \citep[see, e.g., reviews by][and references therein]{Schekochihin_2009_review,  Horbury_anisotropy, Oughton_anisotropy_review}. 
\citet{Horbury_2008_anisotropy, Wicks_anisotropy, Kiyani_2012} have shown that when the field/flow angle $\Theta_{BV}$ is $\Theta_{BV} = 90 ^{\circ}$, then the inertial range range scales like either $\alpha_{B} \approx \textminus 5/3$, or sometimes $\approx \textminus 3/2$, consistent with a critical balance cascade and dynamical alignment models respectively. In the parallel direction, $\Theta_{BV} = 0 ^{\circ}$, it is nearer $\alpha_{B} \approx-2$. In contrast, when a global magnetic field is utilized to estimate $\theta_{BV}$, no anisotropy in the spectral index as a function of $\Theta_{BV}$ is observed \citep{Tessein_2009_no_anisotropy, Chen_2011_no_anis_with_global}. Though it is not shown here, we find no correlation between $\Theta_{BV}$ and $\alpha_{B}$, using a global magnetic field. A similar result was obtained when considering the dependence of $\alpha_{B}$ on plasma $\beta$, suggesting that these two parameters are not related to the steepening of the spectrum. Further work to clarify the debate between a local, scale-dependent and global background magnetic field and it's role on the spectral evolution is presented in a companion paper (Sioulas et al, Submitted).

Our findings will help us gain a better understanding of how solar wind turbulence is generated and transported and will guide future models of solar wind turbulence.
\begin{acknowledgments}

This research was funded in part by the FIELDS experiment
on the Parker Solar Probe spacecraft, designed and developed under NASA contract
NNN06AA01C; the NASA Parker Solar Probe Observatory Scientist
grant NNX15AF34G and the  HERMES DRIVE NASA Science Center grant No. 80NSSC20K0604. 
The instruments of PSP were designed and developed under NASA contract NNN06AA01C. 

\end{acknowledgments}

\begin{appendix}


\begin{figure*}[htb!]
\centering
\setlength\fboxsep{0pt}
\setlength\fboxrule{0.0pt}

\includegraphics[width=1\textwidth]{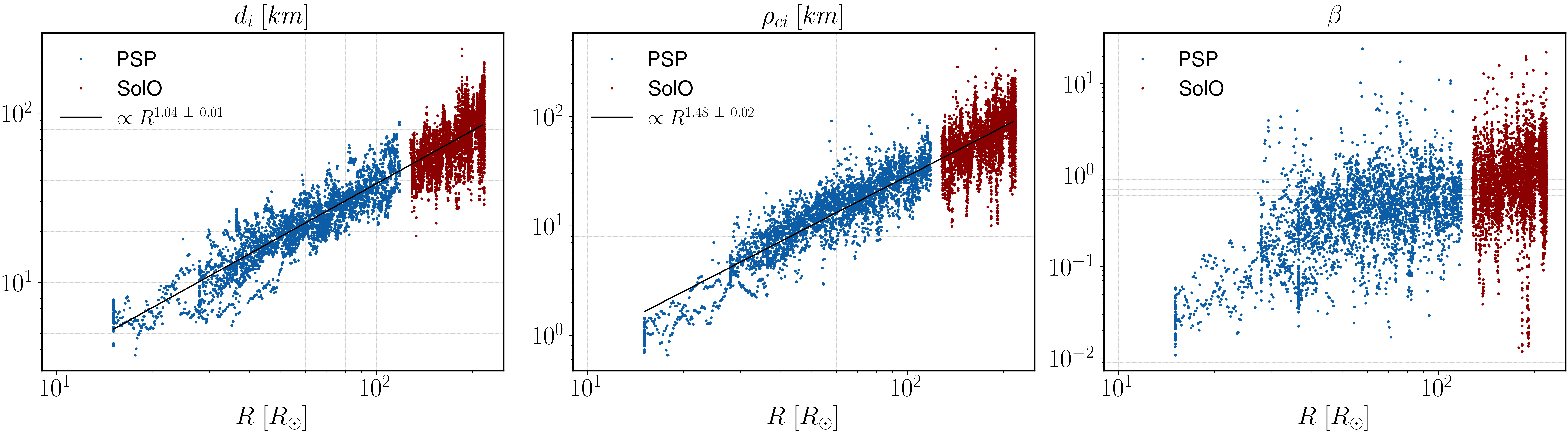}
\caption{\textcolor{blue}The radial evolution of (a) the ion inertial length 
 $d_{i}$, (b) the ion gyroradius$\rho_{i}$, (c) ion plasma $\beta$}\label{fig:overview}
\end{figure*}

\section{\label{sec:Appendix}\\ Data Selection and Processing}

We analyze magnetic field data from the Flux Gate Magnetometer (FGM) \citep{bale_fields_2016}, as well as plasma moment data from the Solar Probe Cup (SPC) and Solar Probe Analyzer (SPAN) part of the Solar Wind Electron, Alpha and Proton (SWEAP) suite between January 1, 2018, and June 15, 2022, encompassing the first twelve perihelia (E1-E12) of the PSP mission. SPC data have been utilized for E1-E8, while SPAN data for E9-E12. Quasi-thermal noise derived data \citep{Moncuquet_2020}, have been preferred over SPAN or SPC data when available.  Additionally, magnetic field and particle moment measurements from the Magnetometer (MAG) instrument \citep{horbury_solar_2020}, prioritizing burst data when available, and the Proton and Alpha Particle Sensor (SWA-PAS) \citep{owen_solar_2020} onboard the SO mission between June 1, 2018, to March 1, 2022, were considered. Following consideration of quality flags, time intervals that were found to be missing $ \geq 1 \%$ and/or $ \geq 10 \%$, in the magnetic field and particle timeseries have been omitted from further analysis. The remaining intervals have been resampled linearly to the highest cadence possible, based on their initial resolution. Finally, in order to eliminate spurious spikes, a Hampel filter \citep{davies_identification_1993} was applied to the plasma time series. \par
 
Converting the spacecraft-frame frequency derived PSD, $F(f_{sc})$ to a wavenumber PSD, $E(\kappa)$, far from the sun is possible by means of Taylor's hypothesis (TH) \citep{taylor_spectrum_1938}, $ \kappa  = {2 \pi f_{sc}}/{V_{SW}}$, that becomes questionable when both the Alfv\'en and spacecraft velocity are comparable to the velocity of the solar wind. Therefore, a modified version of Taylor's hypothesis that accounts for both wave propagation and spacecraft velocity is adopted \citep{Klein_2015}: in the above expression for $\kappa$ $V_{sw}$ is replaced by  $V_{tot} =| \boldsymbol{V}_{sw} + \boldsymbol{V}_{a} - \boldsymbol{V}_{sc}|$  where $\boldsymbol{V}_{sc}$ is the spacecraft velocity, where turbulence is assumed to be dominated by outwardly propagating Alfv\'en waves. Note that the TH remained either moderately or highly valid for the majority of time intervals examined. with only $\sim 1.53 \%$ of the intervals under study exhibiting $M_{A} < 1.5$, including a number sub-Alfv\'enic intervals during  PSP $E_{8}-E_{12}$   ($\sim 0.45 \%$ of the entire dataset).  Figure \ref{fig:overview} illustrates the radial evolution of ion inertial length $d_{i}$, ion gyroradius $\rho_{i}$, and ion plasma $\beta$, quantities relevant to this study.
\end{appendix}

\bibliography{main}{}
\bibliographystyle{aasjournal}



\end{CJK*}
\end{document}